\documentclass[groupedaddress, superscriptaddress, amsmath, amssymb,twocolumn]{revtex4-2}
\usepackage{graphicx, subfigure}
\usepackage{bm}
\usepackage{physics}

\begin{document}

\title{Electronic resonances in expanding non-neutral ultracold plasma}

\author{S. Ya. Bronin}
\affiliation{Joint Institute for High Temperatures of the Russian Academy of Sciences, Izhorskaya St. 13, Bldg. 2, Moscow 125412, Russia}
\author{E. V. Vikhrov}
\affiliation{Joint Institute for High Temperatures of the Russian Academy of Sciences, Izhorskaya St. 13, Bldg. 2, Moscow 125412, Russia}
\author{S. A. Saakyan}
\affiliation{Joint Institute for High Temperatures of the Russian Academy of Sciences, Izhorskaya St. 13, Bldg. 2, Moscow 125412, Russia}
\affiliation{National Research University Higher School of Economics (NRU HSE), Myasnitskaya St. 20, Moscow 101000, Russia}
\author{B. B. Zelener}
\affiliation{Joint Institute for High Temperatures of the Russian Academy of Sciences, Izhorskaya St. 13, Bldg. 2, Moscow 125412, Russia}
\author{B. V. Zelener}
\email{bzelener@mail.ru}
\affiliation{Joint Institute for High Temperatures of the Russian Academy of Sciences, Izhorskaya St. 13, Bldg. 2, Moscow 125412, Russia}
\date{\today}

\begin{abstract}
    Calculations of the spectrum of eigen oscillations of an inhomogeneous ultracold plasma are presented both in the absence and in the presence of charge
     imbalance. The collective modes of these plasma oscillations are recorded in experiments as absorption resonances of a radio frequency electric field.
      It is shown that in the presence of friction of the electronic component, a discrete spectrum of plasma eigen oscillations is formed. 
      The dependence of the frequency of these resonances on the plasma expansion time in the presence of a charge imbalance was obtained.
       There is good agreement with the experiments of different authors.

       \end{abstract}

\maketitle

\section{Introduction}
        Experimental study of ultracold plasma (UCP) makes it possible to find the
    basic physical mechanisms of the processes occurring in any fully ionized plasma, both in case
    of stationary conditions and during plasma expansion. This is due to the well-controlled initial
    conditions and relatively slow expansion dynamics. In addition, only the Coulomb interaction
    between particles is essential in the UCP, while other interactions can be neglected.
    At present, a fairly large amount of experimental data have been obtained for the UCP of
    various chemical elements (Xe, Sr, Rb, Ca) with various densities, various numbers of particles,
    and various initial temperatures of electrons and ions. Many theoretical papers devoted to this
    field have also been published (see reviews \cite{killian2007ultracold, lyon2016ultracold}).
    Most of the experimental results for the UCP are associated with the study of ions, for which
    various diagnostic methods, including optical methods, have been developed. For the study of
    electrons, diagnostic methods are associated with the use of constant and alternating electric
    fields. In \cite{killian1999creation, twedt2010electron}, the process of electron evaporation was studied 
    by means of constant electric
    field in the xenon UCP, that is, the escape of a part of the electrons from the plasma cloud.
   
    Results were obtained for the fraction of remaining electrons as a function of the number of
    particles, density and the initial kinetic energy of the electrons.
    In \cite{kulin2000plasma}, in addition to a constant field, a radio-frequency field with frequency of 5 to 40 MHz and
    with wavelength much larger than the plasma size was also used to study electrons in the UCP of
    xenon atoms. By means of scanning the frequency of the rf field within the said range at various
    times of the plasma expansion, the authors of \cite{kulin2000plasma} recorded the resonant signal of the electron
    output at a certain frequency. Authors of \cite{fletcher2006observation} observed in the xenon UCP collective modes which
    were excited by radio-frequency electric fields and were detected by means of enhanced electron
    emission during plasma expansion. In \cite{lyon2016ultracold, kulin2000plasma, fletcher2006observation,
     twedt2012electronic, wilson2013density},
    dependences of the resonant frequencies of
    electrons on the charge imbalance in the UCP were obtained. Analysis of the results of
    experiments by means of an rf electric field requires a study of eigen oscillations of an
    inhomogeneous non-neutral plasma. The first studies of this kind \cite{trivelpiece1959space, dattner1963resonance,
     gaigneaux1986temperature} 
    are devoted to bounded
    plasmas with cylindrical symmetry, where the electron component is confined by a strong
    magnetic field. A detailed description of these studies is given in the monograph \cite{davidson2001physics}.
    Papers \cite{lyubonko2010energy, bergeson2003neutral} are devoted to the study of eigen oscillations of a spherically
     symmetric bounded
    plasma, in which the electronic component is held by an electric field caused by a charge imbalance.
    As shown in \cite{bergeson2003neutral}, in the absence of friction, i.e. momentum transfer from the electronic component
    to the ion one, the spectrum of eigen vibrations is continuous. 
      In \cite{lyubonko2010energy} the levels of the fundamental mode of plasma eigen oscillations at
       the beginning of expansion ($t=0$) were obtained depending on the charge imbalance. The interaction of plasma
        with a radio frequency field was simulated using molecular dynamics method. However, it was not possible
         to describe the dependence of the spectrum on the imbalance and plasma density during plasma expansion, 
      as was observed in experiments \cite{kulin2000plasma,fletcher2006observation}.

        In this paper we use a method based on solving the known equation for the electric field. This takes into account
         the interaction of the field with the
    plasma by means of conductivity $\mu(\omega, {\bf r})$ and dielectric permittivity $\varepsilon(\omega, {\bf r})$
    ($\omega$ is frequency of the
    rf electric field), which relate the local values of the current density and electric induction to the
    local value of the electric field intensity. Such statement of the problem limits the range of its
    applicability to conditions assuming that the mean free path of electron is small compared to the
    characteristic length of the field variation. The calculations were performed in relation to the experimental
     conditions \cite{kulin2000plasma, fletcher2006observation}. 
    The use of this
    method allowed us to explain the dependence of the spectrum of eigen oscillations of a spherical
    plasma on the charge imbalance and on the expansion time. At the same time, good agreement with
    experimental data was obtained \cite{kulin2000plasma, fletcher2006observation}.

\section{Calculation model}

        The calculations are performed for the experimental conditions of \cite{kulin2000plasma, 
        fletcher2006observation}. 
    We consider the plasma
    that emerges upon single ionization of a limited volume of a cooled neutral gas ($\sim1$~mK).
    Characteristic initial dimensions of the plasma $\sigma_0$ in experiments are limited to fractions of a
    centimeter, the number of ions and electrons $N_i$, $N_e$ are from $10^4$ to $10^8$ and the temperature $T_e$
    of electrons formed during the ionization is between one and few hundred degrees Kelvin. At the
    initial stages of the free expansion of the plasma over times of the order of $\sigma_0 / v_{T_e}$ ($v_{T_e}$ is the
    thermal speed of electrons), that is, of the order of fractions of a microsecond, fast electrons
    leave the plasma. The resulting charge imbalance creates an electric field that prevents further
    leakage of electrons, therefore $\Delta N = N_i - N_e$ remains constant. In order to calculate $\Delta N$, the 
    results of \cite{killian1999creation, vikhrov2021ion} are used. Plasma expansion takes place at a rate characteristic for the ion component,
    which are several orders of magnitude lower than that for the electron component. This makes it
    possible to consider the configuration of the ion component as quasi-stationary when studying \cite{vikhrov2021ion}
    the motion of the electronic component of the plasma. In particular, in consideration of the
    motion of the electronic component, the current value of the plasma size $\sigma(t) (\sigma(0) = \sigma_0)$ is
    assumed to be constant. This said size is determined by the equality $\sigma_0(t) = \sqrt{\langle r^2 / 3 \rangle}$ 
    where angular brackets denote averaging over the ionic configuration. We consider interaction of the
    electronic component with an electromagnetic field whose wavelength is much larger than the
    plasma size, which makes it possible to neglect the influence of the vector potential.

        In the linear approximation for potential $\Phi$, the electric field ${\bf E} = -\nabla\Phi exp(i\omega t)$
    arising under the action of an external field ${\bf E}_0 = {\bf k}E_0 cos(\omega t)$ (the field corresponds 
    to the orbital
    number $l=1$; ${\bf k}$ is the unit vector along the z axis) is determined by the following equation:
    \begin{equation}
        \nabla(\varepsilon \nabla\Phi) = 0.
        \label{eq_one}
    \end{equation}
    
      Here the permittivity is related to the conductivity $\mu$ as follows \cite{landau2013electrodynamics} : 
         
    \begin{equation}
        \begin{gathered}
            \varepsilon = 1-\frac{4\pi i}{\omega}\mu = \varepsilon_1 + i\varepsilon_2\\
            \mu = \mu_1 + i\mu_2 \sim n_e(r),
        \end{gathered}
    \end{equation}
    where $n_e(r)$ is the electron density.
    
    Setting $\Phi = (\varphi_1 + i\varphi_2)\exp(i\omega t)$  into equation (1), we obtain 
     a system of second-order linear equations (\ref{A7}) for
    the functions $\varphi_1$ and $\varphi_2$,
     with two boundary conditions in the center and two at infinity(the complete derivation of the system of equations (\ref{A7}) is presented in Appendix \ref{sct_A1}):

    $\varphi_1(0) = \varphi_2(0) = 0$ at $r = 0$, $\varphi_1 + E_0r \rightarrow 0$, $\varphi_2 \rightarrow 0$
    at $r \rightarrow \infty$.

        In our calculations, the region where the electron concentraion differs from zero is limited. 
    At $r < x_0\sigma$, $n_e \sim exp(-r^2 / 2\sigma^2)$; at $x_0\sigma < r < R = (x_0 + 1/x_0)\sigma$, concentration varies linearly
    from $n_e(x_0\sigma)$ to zero while remaining smooth at the point $x_0\sigma$. 
       The fact that the concentration in the 
    region $r > R$ is zero makes it possible to replace the boundary conditions at $r \rightarrow \infty$ by equivalent
    conditions at $r = R$:
    \begin{equation}
        \begin{gathered}
            2\varphi_1 + R \varphi_1^{\prime} = -3E_0\\
            2\varphi_2 + R \varphi_2^{\prime} = 0.
        \end{gathered}
    \end{equation}

        The chosen scheme for numerical calculation of equations (\ref{A7}) consists in selecting $\varphi_1^{\prime}(0)$ and
        $\varphi_2^{\prime}(0)$ which would fulfill the boundary conditions at $r = R$. With this calculation scheme, the
    admissible values of $x_0$ are determined by the accuracy of the computing device employed, since
    the accuracy of determining $\varphi_1^{\prime}(0)$ and $\varphi_2^{\prime}(0)$ is limited by the number of binary 
    digits in the representation of numbers. In our case, the allowed values of $x_0$ should not exceed $x_0 = 3$.
   
    Numerical solution of equations (A7) (Appendix A) makes it possible to determine the conditions for an increase of heat release in the plasma, which is accompanied in the experiment by an increase in the registered flux of electrons leaving the plasma region.
     The total heat release is given by the integral (B5) (see Appendix \ref{sct_A2}). 
    As the frequency of the external electric field approaches the frequency of eigen oscillations, the heat release tends to infinity, which makes it possible to determine the spectrum of eigen oscillations of the plasma under study.

\section{Absorption resonances at $\Delta N = 0$}

  First, let's consider the case of a small imbalance ($\Delta N = 0$), when the electric field generated by it can be neglected. In this case, 
the frequency dependence of conductivity and permittivity is given by the known relations \cite{landau2013electrodynamics}:

               \begin{equation}
        \begin{gathered}
            \mu = \frac{e^2n_e(r)}{m_e}\frac{\omega}{\omega\nu + i\omega^2}\\
            \varepsilon = 1 - \frac{\omega_p^2(r)}{\omega^2 - i\omega\nu},
        \end{gathered}
        \label{eq_Cond}
    \end{equation}
    where $\nu$ is frequency of electron-ion collisions and $\omega_p(r) = \sqrt{4\pi e^2 n_e(r) / m_e}$. 
    The value of $\nu$ is calculated by means of the formula for a weakly coupled plasma \cite{spitzer2006physics}:
    \begin{equation}
        \begin{gathered}
            \nu = \frac{4\sqrt{2\pi}n_ie^2ln\Lambda}{3\sqrt{m_eT_e^3}}\\
            ln\Lambda = ln\frac{1}{\sqrt{3\Gamma_e^3}}\\
            \Gamma_e = \frac{e^2\sqrt[3]{4\pi n_e / 3}}{T_e},
        \end{gathered}
        \label{eq_Freq}
    \end{equation}
    where $T_e$ is equal to the average electron temperature in the plasma region and $n_i$ is ionic concentration.
    The condition for the applicability of (\ref{eq_Cond}) is the smallness of the free path of electrons in
    comparison with the characteristic length of the electric field inhomogeneity $d$:
    \begin{equation}
        d \gg v_{T_e} / v,
    \end{equation}
    where $v_{T_e}$ is thermal electron velocity.

\begin{figure}[ht!]
    \includegraphics[width=0.9\linewidth]{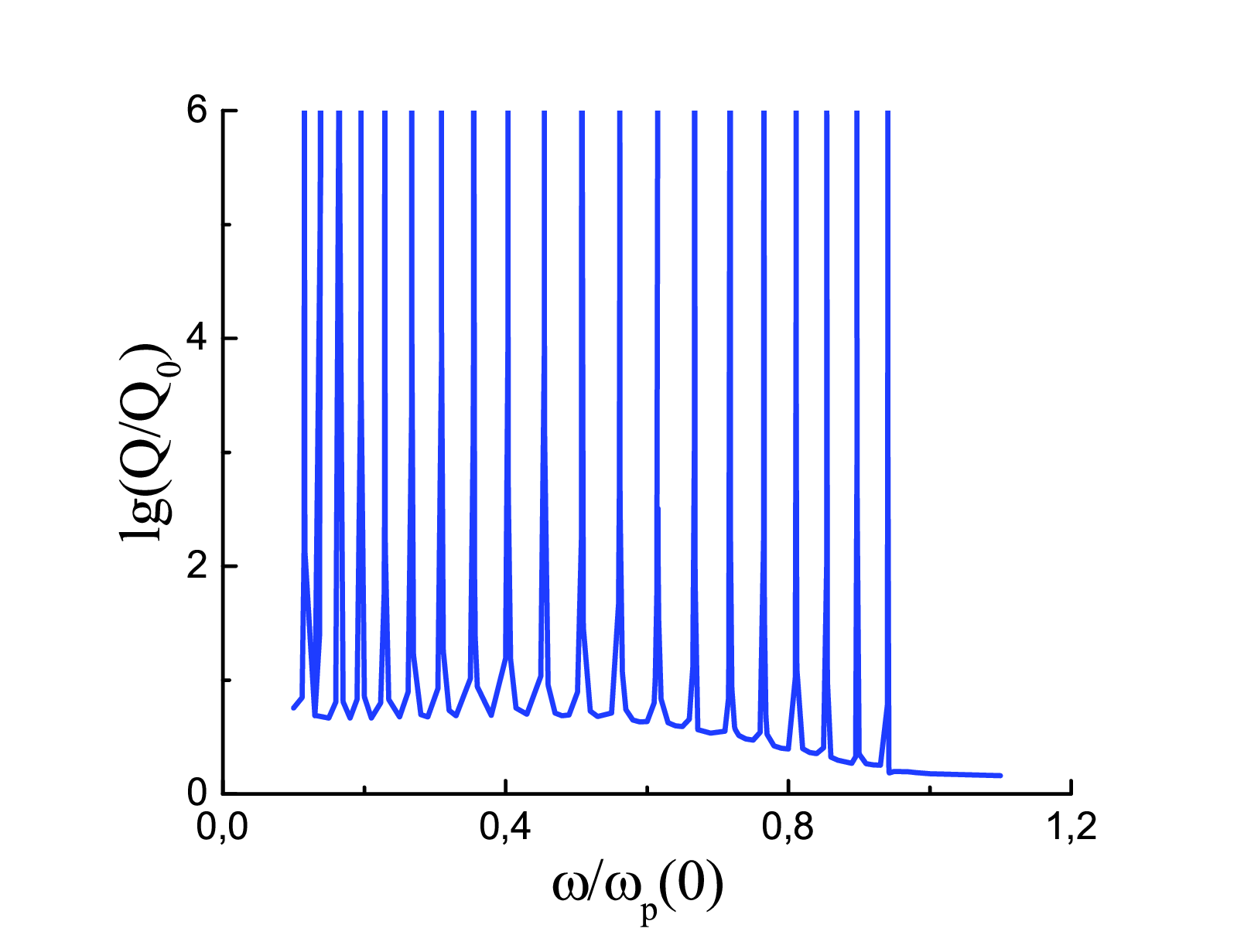}
    \caption{Dependence logarithm of dimensionless $Q / Q_0$ on $\omega / \omega_p(0)$ at $\nu = 0.4\omega_p(0)$, $x_0 = 3$.}
    \label{fig_1}
\end{figure}

The solution of equations (A7) with boundary conditions (3) describe the plasma response
 to the effect of external rf radiation with a frequency of $\omega$ and an amplitude of $E_0$.
  Figure 1 shows the dependence of heat release $Q/Q_0$ where $Q_0=E(0)^2\int\mu_1 d\bm{r}/2 $ 
   on $\omega / \omega_p(0)$ at $\nu = 0.4\omega_p(0)$ . Calculations demonstrate 
  distinct heat release resonances, similar to Tonks-Dattner resonances. The same resonances of 
  heat release were observed in calculations \cite{lyubonko2010energy} performed by the method
   of molecular dynamics. When the resonant frequency $\widetilde{\omega}$ is approached, 
   the heat release grows inversely
    proportional to the square of the distance from the resonance $Q\sim 1/ (\omega - \widetilde{\omega})^2$.

The solutions of equations (A7) $\varphi_1$ and $\varphi_2$ also
 increase indefinitely inversely $\omega - \widetilde{\omega}$, 
 and the limits of the products $(\omega -\widetilde{\omega})\varphi_{1,2}$ 
 for $\omega \to\widetilde{\omega}$ equal to $\widetilde{\varphi}_{1,2}$   
  are solutions of homogeneous equations (A7) 
 (with $E_0=0$), that is, they represent the eigen oscillations of the plasma
  formation under consideration. In fact, the lifetime of real oscillations 
  with a resonant frequency is limited by the condition of quasi-stationarity 
  and does not exceed the interval $\Delta t$, within which the ion distribution can be
   considered unchanged. Accordingly, the width of the observed resonance
    cannot be less than $\Delta{\omega}=2\pi/\Delta t$. The frequencies of such states 
    form a discrete spectrum containing an apparently infinite number of frequencies 
    in the region $\omega<\omega_p (0)$.

\begin{figure}[ht!]
    \includegraphics[width=0.9\linewidth]{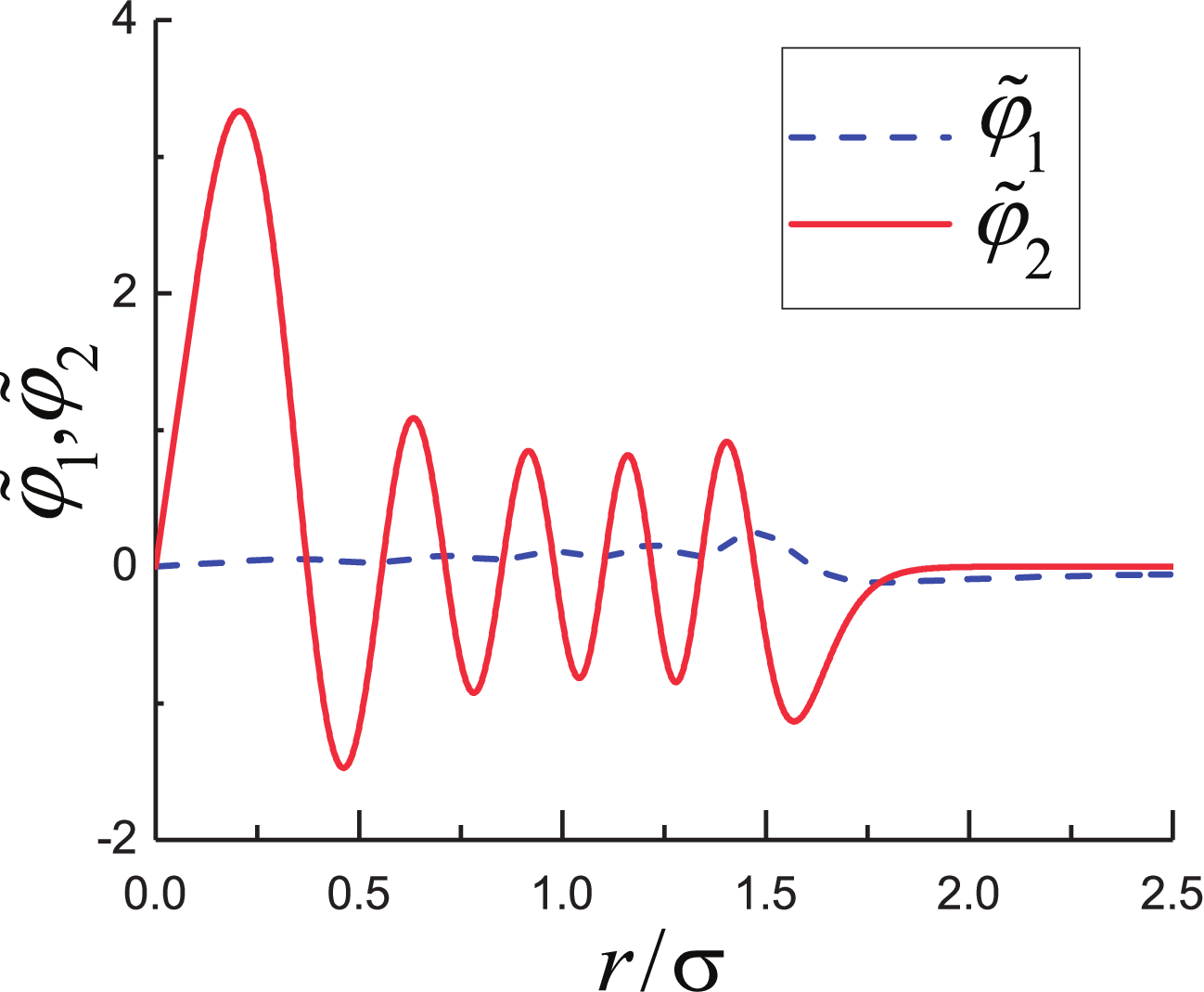}
    \caption{Normalized eigensolutions of homogeneous equations (\ref{eq_Motion}) corresponding to the eigen
        frequency $\omega/\omega_p(0) = 0.51$ for $\omega_0 = 0$, $\nu = 0.4\omega_p(0)$, $x_0 = 3$ depending on 
        $r / \sigma$: $\widetilde{\varphi}_1$ is the blue curve, $\widetilde{\varphi}_2$ is the red curve.}
    \label{fig_2}

\end{figure}
  
    Figure \ref{fig_2} shows an example of a solutions of homogeneous equations (\ref{A7}) normalized by means 
    of the equality $\int (\widetilde{\varphi}_1^2 + \widetilde{\varphi}_2^2)r^2dr / \sigma^3 = 1$ corresponding to the tenth resonance 
    $(\widetilde{\omega} = 0.51 ... \omega_p(0))$. The blue
    curve corresponds to the function $\widetilde{\varphi}_1$, the red curve corresponds to the function $\widetilde{\varphi}_2$.
       The number of extrema of the function ${\widetilde\varphi_2}$ is equal to the serial number $N$ of resonances,
    numbered from right to left. Choosing $\sigma / N $ as the characteristic size of the field inhomogeneity,
    we obtain from the inequality (\ref{eq_Freq}), which determines the range of applicability of the equations
    considered, the following estimate for the maximum number of eigen frequencies:
    \begin{equation}
        N < \sigma v / v_{T_e}.
        \label{eq_Freq_Num}
    \end{equation}

\begin{figure}[ht!]
        \includegraphics[width=0.9\linewidth]{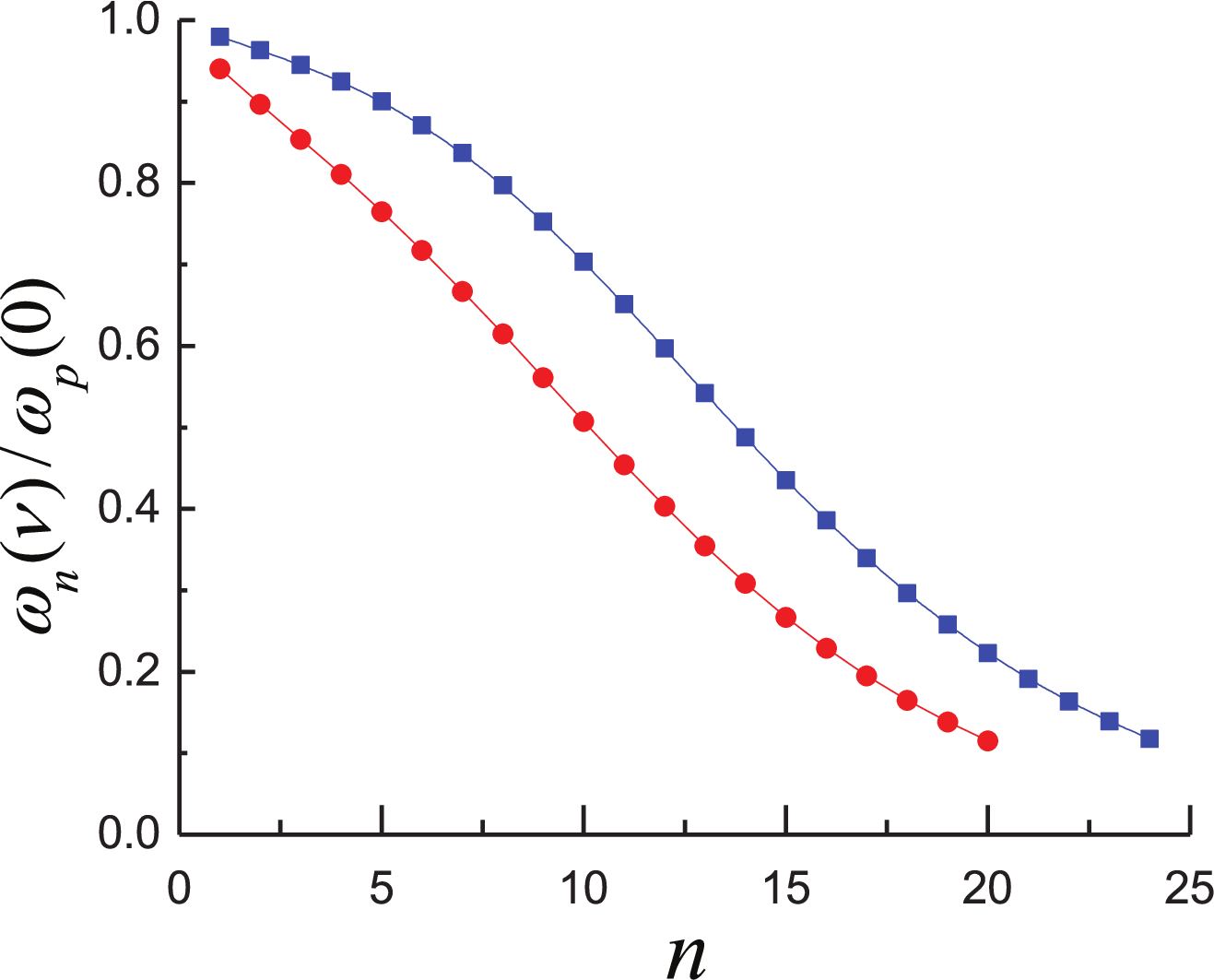}
        \caption{eigen frequencies $\omega_n(\nu)/\omega_p(0)$, where $n$ is the serial number of the resonance, for $\omega_0 = 0$ and for two values of the friction 
        coefficient: $\nu / \omega_p(0) = 0.1$ corresponds to the blue curve and $0.4$ corresponds to the red curve.}
        \label{fig_3}
    \end{figure}

    With $\nu / \omega_p(0)$ decreasing, the distance between eigen frequencies decreases. Figure \ref{fig_3} shows
    eigen frequencies $\omega_n(\nu)$, numbered starting from the maximum frequency $\omega_1 \approx \omega_p$, for two values of the
    friction coefficient:  $\nu / \omega_p(0) = 0.4$ is for the blue dash curve and $\nu / \omega_p(0) = 0.1$ is for the red solid
    curve.
        
        For small friction coefficients, the calculations are impossible due to the same reasons as for large
    values of $x_0$. Calculations for $\nu / \omega p(0)$ values greater than $0.1$ show that the distance between
    resonant frequencies decreases with decreasing friction coefficient which agrees with the well-known fact that in the absence of friction the spectrum is continuous \cite{bergeson2003neutral}, similarly to the
    Trivelpiece-Gould spectrum \cite{dattner1963resonance}.

\section {Calculation of absorption resonances with account taken of the charge imbalance}
        In order to calculate the spectrum of eigen oscillations of an inhomogeneous plasma, with account
    taken of the charge imbalance, it is necessary to determine the conductivity and dielectric
    permittivity of the plasma in the presence of a constant electric field. A similar problem was
    solved in \cite{dattner1963resonance} for the conditions where the electronic component is confined by means of a
    magnetic field. After a quasi-stationary mode for the electric field is established in the presence
    of spherical symmetry, the following self-similar expression is valid \cite{vikhrov2021ion}:
    \begin{equation}
        \bm{E}_{st}(\bm{r},t) = \frac{e \Delta N}{\sigma^3(t)}\bm{r}\eta\left(\frac{r}{\sigma(t)}\right),
    \end{equation}
    where $\eta(\rho)$ is constant in the inner part of the plasma and, decreases proportionally to $1 / \rho^3$
    outside it. For the potential energy of electrons, we have:
    \begin{equation}
        \label{eq_El_Potential}
        \begin{gathered}
            {e}\varphi_{st} = \frac{e^2\Delta N}{\sigma^3(t)}\int\limits_0^r r^{\prime} \eta(r^{\prime} / \sigma(t))dr^{\prime}\\ 
            \approx\frac{e^2\Delta N}{\sigma^3(t)}\eta(0)\frac{r^2}{2}= \frac{m_e\omega_0^2(t)}{2}r^2.
        \end{gathered}
    \end{equation}
    With account taken of the following estimates: $\eta(0)\approx 1$; $\Delta N \approx \sqrt{kT_eN_i\sigma_0} / e$ 
    \cite{killian1999creation, vikhrov2021ion}, we obtain:
    \begin{equation}
        \omega_0^2(t)\approx\frac{e\sqrt{kT_eN_i\sigma_0}}{m_e\sigma^3(t)}.
    \end{equation}

        Let us represent the solution of the equation of motion of an individual electron in an external
    electric field ${\bf E} = {\bf E}({\bf r}) exp(i\omega t)$, with account taken of the friction coefficient $\nu$:
    \begin{equation}
        m_e\ddot{\bm{r}} + m_e\bm{r}\omega_0^2 + m_e\nu\dot{\bm{r}} = -e\bm{E}(\bm{r})e^{i\omega t}
        \label{eq_Motion}
    \end{equation}
    in following form:
    \begin{equation}
      \bm{r}(t) = \bm{r}_0(t) + \Delta\bm{r}(t)
    \end{equation}
    where $\bm{r}_0(t)$ is the solution of this equation for $\bm E(\bm{r}) = 0$ and
    \begin{equation}
        \Delta \bm{r} (t) = -\frac{e}{m_e\omega_1}\int\limits_0^t e^{-\nu(t - t^{\prime})/ 2}
        sin(\omega_1(t - t^{\prime}))\bm E (\bm{r}(t^{\prime}))e^{i\omega t^{\prime}}dt^{\prime}
    \end{equation}
    where $\omega_1^2 = \omega_0^2 - \nu^2 / 4$.

        In the conditions where the mean free path of electron ($v_{T_e}$ is the thermal velocity of electrons) is
    less than the characteristic size of the field variation $d$ (inequality (6)), the following 
    expression is valid for $\Delta \bm{\dot{r}}  (t)$:
    \begin{equation}
        \begin{gathered}
            \Delta \bm{\dot{r}}  (t) \approx \frac{e}{m_e\omega_1}\bm E\left(\bm{r} (t)\right) e^{i\omega t}\\
            \cdot \int \limits_0^{\infty} e^{-(\nu / 2 + i\omega t)t}\cdot\left(\omega_1 cos\omega_1 t - 0.5\nu sin\omega_1 t\right)dt\\
            = -\frac{e\omega}{m_e}\bm{E} \left(\bm{r} (t)\right)e^{i\omega t} \frac{\omega}{\omega\nu - i(\omega_0^2 - \omega^2)}.
        \end{gathered}
    \end{equation}

        For the electron current we have:
    \begin{equation}
        \label{eq_Electron_Current}
        \bm j(\bm{r}) = \int f(\bm{r}(0), \dot{\bm{r}}(0))\bm{\dot{r}}  (t)d\bm{r}(0)d\dot{\bm{r}}(0), 
    \end{equation}
    where $f(\bm{r}(0), \dot{\bm{r}}(0))$ is the quasi-equilibrium electron distribution function $\int f(\bm{r}, \dot{\bm{r}})d\bm{\dot{r}} = n_e (\bm{r})$,
    for which the following equality is valid within the linear approximation: $f(\bm{r}(0), \dot{\bm{r}}(0)) = 
    f(\bm{r}(t), \dot{\bm{r}}(t))$.

        After the following change of variables in (\ref{eq_Electron_Current}): $\bm{r}(0), \dot{\bm{r}}(0) \rightarrow \bm{r}(t), \dot{\bm{r}}(t)$,
    we obtain for the current:
    \begin{equation}
        \begin{gathered}
            \bm{j}(\bm{r}) = \mu(\bm{r}, \omega)\bm E(\bm{r}) e^{i\omega t}\\
            \mu(\bm{r}, \omega) = \frac{e^2n_e(\bm{r})}{m_e}\frac{\omega}{\omega\nu - i(\omega_0^2 - \omega^2)} = \mu_1 + i\mu_2,
        \end{gathered}
    \end{equation}
    where $\mu$ is local plasma conductivity.
    For the dielectric permittivity we have: 
    \begin{equation}
        \begin{gathered}
        \varepsilon = 1 - \frac{4\pi i}{\omega}(\mu_1 + i\mu_2)\\
                 =1 - \frac{\omega_p^2}{\omega^2 - \omega_0^2 - i\omega\nu} = \varepsilon_1 + i\varepsilon_2,
        \label{eq_Diel_Pim}
    \end{gathered}
    \end{equation}
    where $\omega_p = \sqrt{4\pi e^2n_e(\bm{r}) / m_e}$ is the plasma frequency.

\begin{figure}
        \subfigure[]{
            \includegraphics[width=0.9\linewidth]{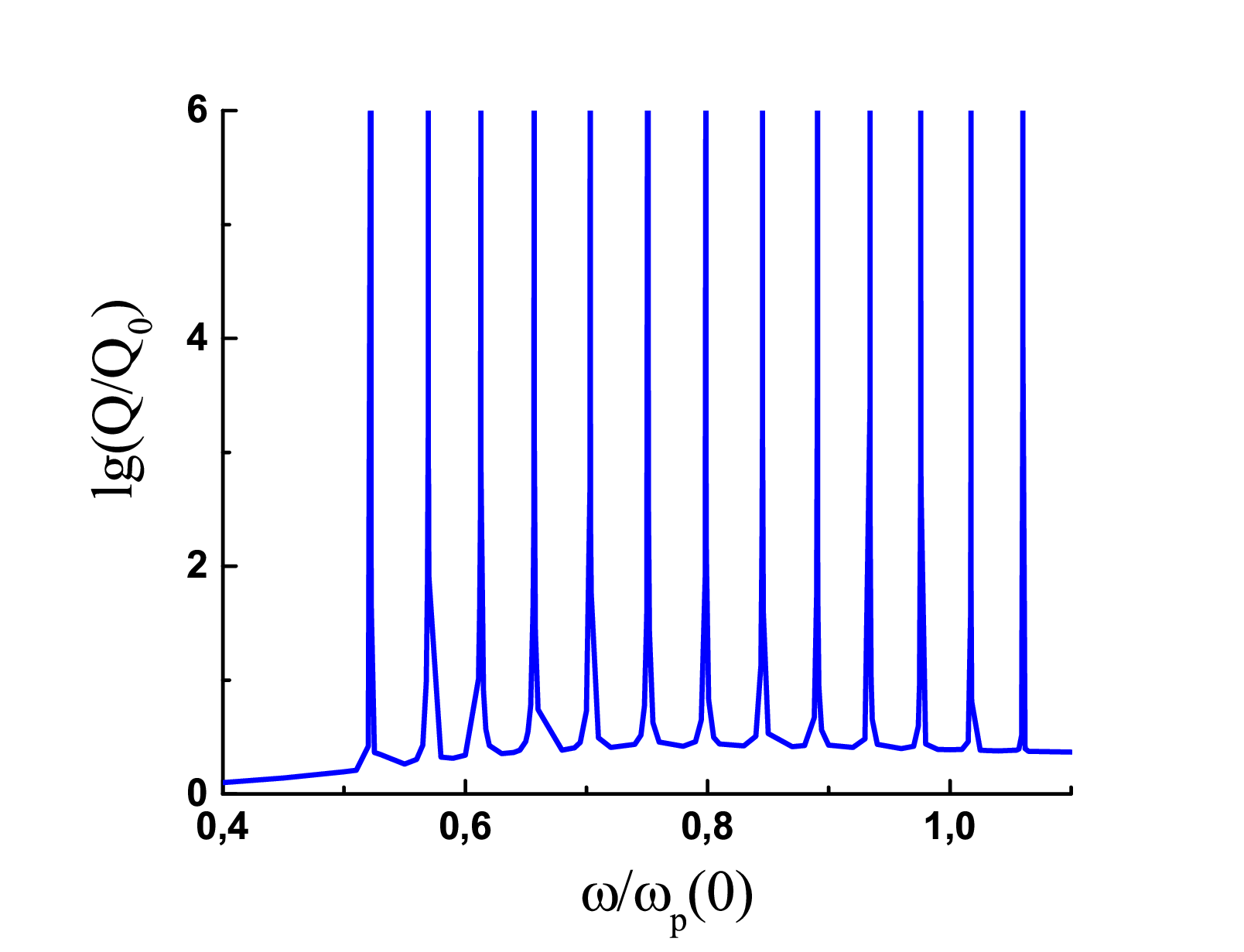}
            \label{fig_5a} 
        }
        \subfigure[]{
            \includegraphics[width=0.9\linewidth]{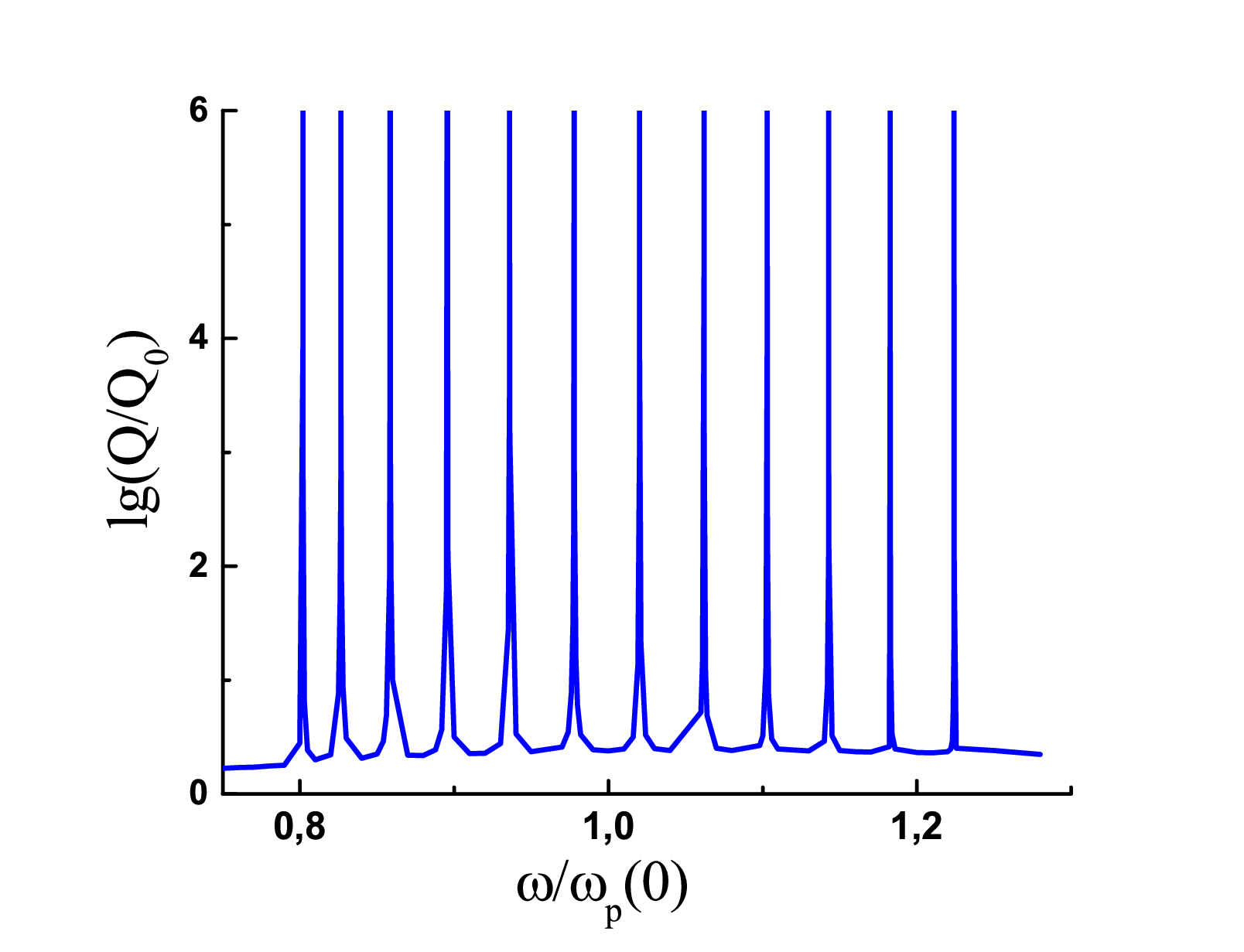}
            \label{fig_5b} 
        }
        \caption{Dependence of the logarithm of the dimensionless quantity $Q/Q_0$ on $\omega/\omega_p(0)$
                at $\nu = 0.4\omega_p(0)$} $x_0 = 3$:
                \subref{fig_5a} $\omega_0 = 0.5\omega_p(0)$;
                \subref{fig_5b} $\omega_0 = 0.8\omega_p(0)$;
        \label{fig_5}
    \end{figure}

        Numerical solution of equations (\ref{A7}) with dielectric permittivity (\ref{eq_Diel_Pim}) 
    show that in the presence of friction, the spectrum remains discrete, but with a finite number of eigen 
    frequencies, limited by the minimal frequency $\sim \omega_0$. Figure \ref{fig_5} shows the dependence of heat 
    release $Q$ on $\omega / \omega_p(0)$ at $\nu=0.4\omega_p(0)$ and $\omega_0 = 0.5\omega_p(0)$ (\ref{fig_5a}) 
    and $\omega_0=0.8\omega_p(0)$ (\ref{fig_5b}) with all eigen frequencies.
        The minimum of the resonant frequency is close to $\omega = \omega_0$. It follows from the above-said
    expression for the frequency $\omega_0$ that its time dependence is determined by the following relation:
    $\omega_0 \sim \sqrt{\Delta N / \sigma(t)^3}$. A similar time dependence of the resonant frequency was observed in
    experiments \cite{twedt2012electronic}. In the stationary expansion mode, where $\Delta N$ and the plasma expansion 
    velocity reach stationary values $(\sigma(t) \sim t)$, the following relation is valid for time dependence of the
    frequency $\omega_0$: $\omega_0 \sim t^{-3/2}$.

\section{Comparison with experiment}

\begin{figure}[ht!]
        \includegraphics[width=0.9\linewidth]{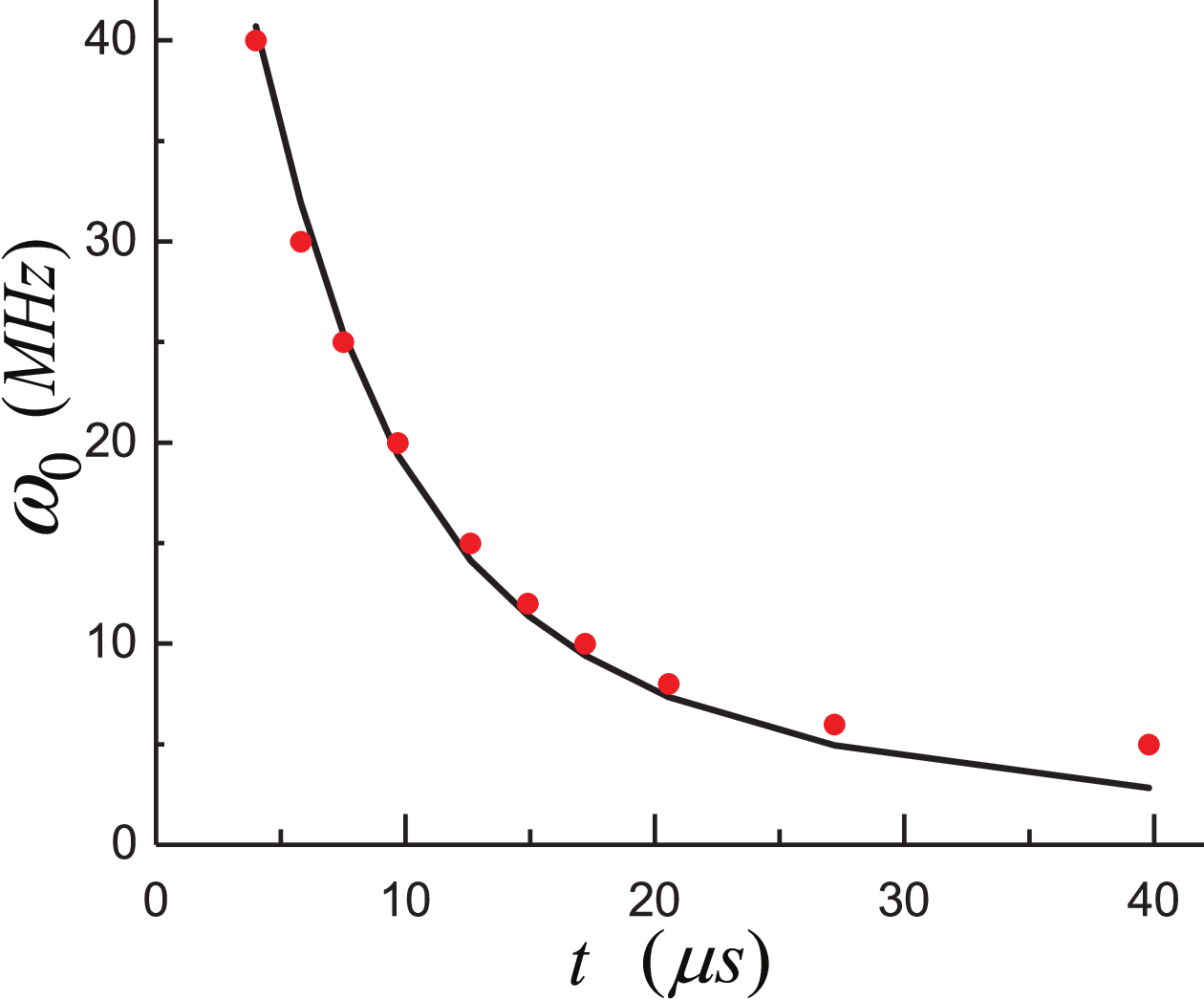}
        \caption{Comparison of calculations with experiment \cite{kulin2000plasma}.Points are for the experiment, the
        curve is for the calculation (\ref{eq_El_Potential}). UCP parameters: $\sigma_0 = 0,022$~cm, $N_i = 8\cdot10^4$,
        $T_{e0} = 26$~K}
        \label {fig_6}
    \end{figure}

        Figure \ref {fig_6} shows the results of our calculation of the
    frequencies of the main resonances $\omega_0$ on the basis of (\ref{eq_El_Potential}) in comparison with the experiment
    \cite{kulin2000plasma}. The ratio $\omega_0(t) / \omega_p(0,t)$ does not depend on time. The experimental values are obtained by
    means of digitizing Fig. \ref{fig_1} of \cite{kulin2000plasma}. Unfortunately, for other parameters of the xenon UCP, \cite{kulin2000plasma}
    does not provide resonance frequencies, but only gives the results of processing for the
    expansion time dependence of the average density.

\begin{figure}[ht!]
        \includegraphics[width=0.9\linewidth]{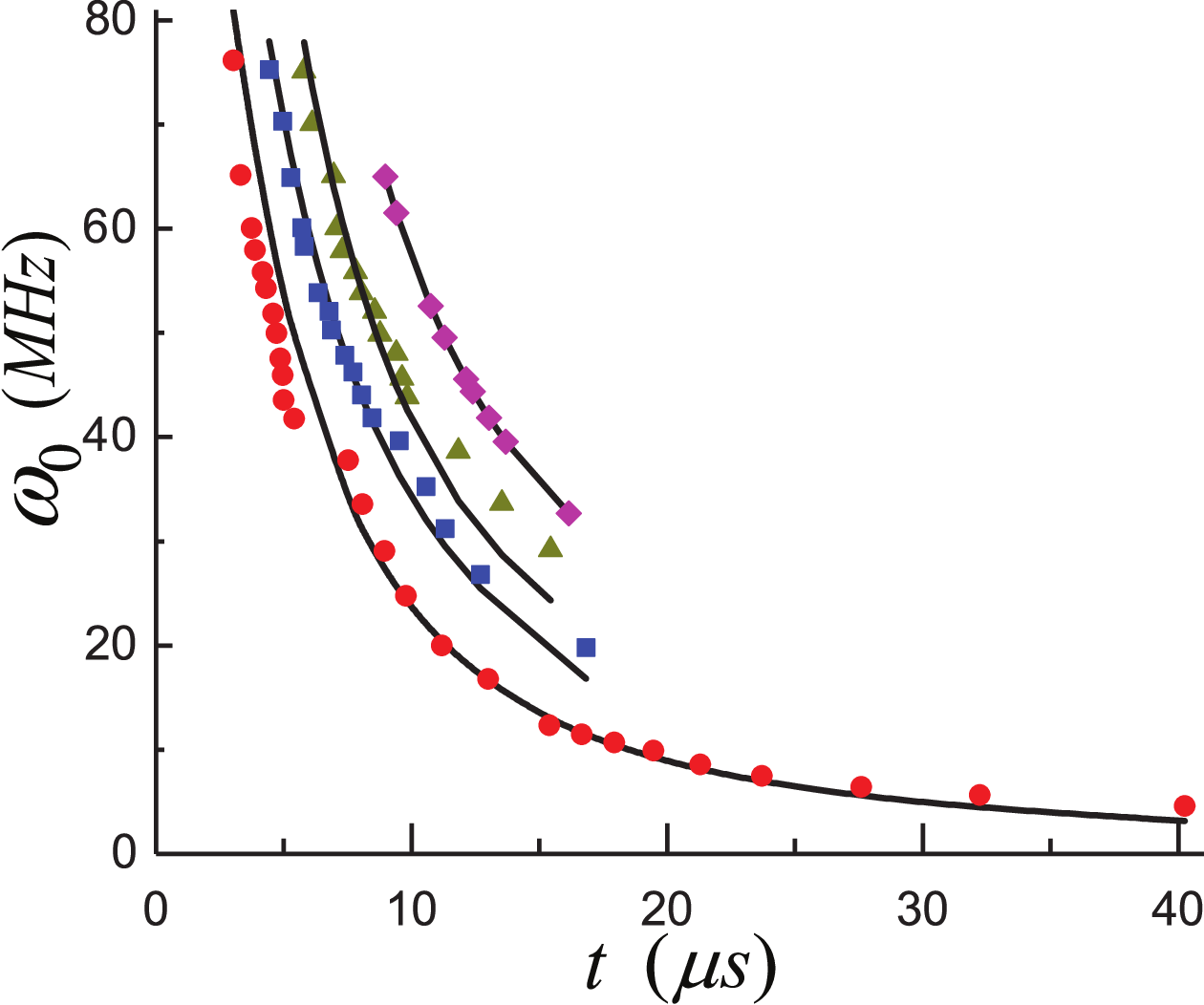}
        \caption{Comparison of calculations with experiment \cite{fletcher2006observation}.
        Points are for the experiment, the lower curve is for calculation (\ref{eq_Diel_Pim}); 
        the next curves are from numerical solution of (\ref{eq_Boundary_Problem})
        with dielectric permittivity (\ref{eq_Diel_Pim}). UCP parameters: $\sigma_0=0.028$~cm, 
        $N_i = 1.5 \cdot 10^6$,
        $T_e=100$~K, $\nu_{ei0} = 24$~MHz.
        }
        \label {fig_7}
    \end{figure}

        Figure \ref{fig_7} shows the results of our calculations of the fundamental frequencies of electronic
    resonances $\omega_0$ and the frequencies of the first three modes on the basis of (\ref{A7}) with dielectric
    permittivity (\ref{eq_Diel_Pim}), in comparison with experiment \cite{fletcher2006observation}. The 
    ratio $\omega_0(t) / \omega_p(0,t)$ in these conditions is
    $0.25$ ($\sigma_0=0.028$~cm, $n_0 = 2 \cdot 10^9$~cm$^{-3}$, $T_e=100$~K). The lower curve (the main resonance $\omega_0$) is
    calculated by means of (\ref{eq_Diel_Pim}). The other three curves are obtained from the above-described
    calculations for the ratios of the next resonances to the main resonance $\omega_k / \omega_0$. The curve
    following the first one corresponds to the 1st experimental resonance from \cite{fletcher2006observation}, the next curve
    corresponds to the 2nd experimental resonance from \cite{fletcher2006observation}, and the last curve corresponds to the 3rd
    experimental resonance from \cite{fletcher2006observation}. Agreement with experiments \cite{kulin2000plasma, fletcher2006observation}
    can be considered good.

\section{Conclusion}
        In the present paper, the spectrum of eigen oscillations of a spherical plasma is
    obtained as a function of charge imbalance. A method is used which is based on solving the
    well-known equations for electric field whose interaction with plasma is taken into account by
    means of dielectric permittivity $\epsilon(\omega,{\bf r})$ (where $\omega$ is frequency of the ref field). The range of
    applicability of this approach is limited by the conditions that assume that the mean free path of
    electrons is small compared to the characteristic length of the field variation. As a result, the main and subsequent modes of resonances in expanding non-neutral ultracold plasma were obtained, which are in good agreement with the available experimental data. Moreover, such calculations are absent in works devoted to this topic. In addition, we were able to show that electronic 
    resonances also arise in the case of a neutral, but not homogeneous, plasma.

\section{Acknowledgments}
        The research was supported by the Russian Science Foundation Grant No. 21-72-00011 in the part
    of processing and analysis of the experiment and No. 23-72-10031 in the part of large-scale
    reworked this article within the framework of the new grant. This work was supported by the
    Ministry of Science and Higher Education of the Russian Federation (State Assignment No. 075-01129-23-00) in the part of creating a program code for the numerical solution of a system of two
    second-order differential equations with complex boundary conditions. This research was also
    supporte by computational resources of HPC facilities at HSE University.

\onecolumngrid
\appendix

\section{} \label {sct_A1}
        Here is a description, omitted in the text, of the transition from equation (\ref{eq_one}) to a system of two
    real equations (\ref{A7}) for $\varphi_1$ and $\varphi_2$
    \begin{equation}
        Re\left[\nabla (\varepsilon\nabla\Phi) e^{i\omega t}\right] = D_1 \cos{(\omega t)} + D_2 \sin{(\omega t)} = 0
    \end{equation}
    
      Equations for $D_1$ and  $D_2$ is obtained by   applying the operator $\nabla$ to the expression in parentheses (A1)    and keeping only the real part:
       \begin{equation}
        \begin{gathered}
            D_1 = \varepsilon_1\Delta\Phi_1 - \varepsilon_2\Delta\Phi_2 + \nabla\varepsilon_1\nabla\Phi_1 - \nabla\varepsilon_2\nabla\Phi_2 = 0\\
            D_2 = -\varepsilon_1\Delta\Phi_2 - \varepsilon_2\Delta\Phi_1 - \nabla\varepsilon_1\nabla\Phi_2 - \nabla\varepsilon_2\nabla\Phi_1 = 0
        \end{gathered}
    \end{equation}

    A system of second-order linear differential equations with coefficients  $\alpha$ and $\beta$ for $\Phi_1$  and  $\Phi_2$  is obtained by combining  $D_1$ and $D_2$ with $\varepsilon_1$ and $\varepsilon_2$:
    \begin{equation}
        \begin{gathered}
            \frac{1}{\varepsilon_1^2 + \varepsilon_2^2}(\varepsilon_1 D_1 - \varepsilon_2 D_2) = \Delta\Phi_1 
            + \frac{\varepsilon_1\nabla\varepsilon_1 + \varepsilon_2\nabla\varepsilon_2}{\varepsilon_1^2 + \varepsilon_2^2}\nabla\Phi_1
            + \frac{\varepsilon_2\nabla\varepsilon_1 - \varepsilon_1\nabla\varepsilon_2}{\varepsilon_1^2 + \varepsilon_2^2}\nabla\Phi_2 = 0\\
            -\frac{1}{\varepsilon_1^2 + \varepsilon_2^2}(\varepsilon_2 D_1 + \varepsilon_1 D_2) = \Delta\Phi_2 
            + \frac{\varepsilon_1\nabla\varepsilon_1 + \varepsilon_2\nabla\varepsilon_2}{\varepsilon_1^2 + \varepsilon_2^2}\nabla\Phi_2
            - \frac{\varepsilon_2\nabla\varepsilon_1 - \varepsilon_1\nabla\varepsilon_2}{\varepsilon_1^2 + \varepsilon_2^2}\nabla\Phi_1 = 0
        \end{gathered}
    \end{equation}
    \begin{equation}
        \begin{gathered}
            \frac{\varepsilon_1\nabla\varepsilon_1 + \varepsilon_2\nabla\varepsilon_2}{\varepsilon_1^2 + \varepsilon_2^2} = \alpha \frac{\bm{r}}{r}\\
            \frac{\varepsilon_2\nabla\varepsilon_1 - \varepsilon_1\nabla\varepsilon_2}{\varepsilon_1^2 + \varepsilon_2^2} = \beta \frac{\bm{r}}{r}\\
        \end{gathered}
    \end{equation}
    \begin{equation}
        \begin{gathered}
            \alpha = \frac{\varepsilon_1\varepsilon_1^{\prime} + \varepsilon_2\varepsilon_2^{\prime}}{\varepsilon_1^2 + \varepsilon_2^2}\\
            \beta = \frac{\varepsilon_2\varepsilon_1^{\prime} - \varepsilon_1\varepsilon_2^{\prime}}{\varepsilon_1^2 + \varepsilon_2^2}
        \end{gathered}
    \end{equation}

    In formulas (A6) $\bm {k}$ - the single vector of the z axis :
    \begin{equation}
        \begin{gathered}
            \Phi_m =\varphi_m(r)\cos(\theta)=\varphi_m\frac{\bm {r}\bm {k}}{r}\\
            \nabla\Phi_m =\frac{\bm {r}}{r}\varphi^{\prime}_m\cos(\theta)+\frac{\bm {k}}{r}\varphi_m-\frac{\bm {r}}{r^2}\varphi_m\cos(\theta)\\
            \frac{\bm {r}}{r}\nabla\Phi_m = \frac{\bm {r}}{r}\varphi^{\prime}_m\cos(\theta)\\
            \Delta\Phi_m = \left(\frac{1}{r^2}\frac{d}{dr}\left(r^2 \frac{d\varphi_m}{dr}\right) - \frac{2}{r^2}\varphi_m\right)\cos(\theta)
        \end{gathered}
    \end{equation}
        
    The final expression for the system of second-order linear differential equations for $\varphi_1$ and  $\varphi_2$    is obtained by using the representation   $\Phi_m$, $\nabla\Phi_m $  and $\Delta\Phi_m$         
            through $\varphi_m$:          
    \begin{equation}\label {A7}
        \begin{gathered}
            \frac{1}{r^2}\frac{d}{dr}\left(r^2 \frac{d\varphi_1}{dr}\right) - \frac{2}{r^2}\varphi_1 + \alpha\frac{d\varphi_1}{dr} + \beta\frac{d\varphi_2}{dr} = 0\\
            \frac{1}{r^2}\frac{d}{dr}\left(r^2 \frac{d\varphi_2}{dr}\right) - \frac{2}{r^2}\varphi_2 + \alpha\frac{d\varphi_2}{dr} - \beta\frac{d\varphi_1}{dr} = 0
        \end{gathered}
    \end{equation}

\section{} \label {sct_A2}
    Derivation of the expression for for total heat release
    (brackets indicate time averaging) $Q = \int d\bm{r} \langle q \rangle$:
    \begin{equation}
        \begin{gathered}
          q = Re\bm{E}Re{\bm{J}} = Re\bm{E}Re(\mu_1 + i\mu_2)\bm{E}
        \end{gathered}
    \end{equation}

    The right side of (B1) is transformed using expressions for ${\bf E} = -\nabla\Phi exp(i\omega t)$   and (16) for $\bm{J}$ :
    \begin{equation}
        \begin{gathered}
          (\mu_1(\nabla\Phi_1\cos(\omega t) -\nabla\Phi_2\sin(\omega t)) - \mu_2(\nabla\Phi_1\sin(\omega t) +\nabla\Phi_2\cos(\omega t)))\cdot\\
          \cdot(\nabla\Phi_1\cos(\omega t) - \nabla\Phi_2\sin(\omega t))  \\
        =  \mu_1((\nabla\Phi_1)^2\cos^2(\omega t) + (\nabla\Phi_2)^2\sin^2(\omega t)) - \mu_2\left[((\nabla\Phi_1)^2 - (\nabla\Phi_2)^2)\frac{\sin(2\omega t)}{2}  
          \nabla\Phi_1\Phi_2 \cos(2\omega t)\right] -\\
           - \mu_1\nabla\Phi_1\Phi_2{\sin(2\omega t)} 
        \end{gathered}
    \end{equation}
    \begin{equation}
        \begin{gathered}
          \langle \cos^2(\omega t)\rangle = \langle \sin^2(\omega t)\rangle = \frac{1}{2}\\
          \langle \cos(2\omega t)\rangle = \langle \sin(2\omega t)\rangle = 0
        \end{gathered}
    \end{equation}

    The expressions for  $\langle q \rangle$    both the function $\varphi_1$ and  $\varphi_2$ - (B4) and for calculating total heat release Q - (B5) are obtained by taking into account the equalities (B3) and  $(\nabla\Phi_m)^2$:
   \begin{equation}
        \begin{gathered}
            \langle q \rangle = \frac{\mu_1}{2}((\nabla\Phi_1)^2 + (\nabla\Phi_2)^2)\\
             (\nabla\Phi_m)^2 = \frac{\varphi_m^2}{r^2} + \cos^2\theta\left(\varphi_m^{\prime 2}
          + \frac{\varphi_m^2}{r^2} - \frac{\varphi_m\varphi_m^{\prime}}{r}\right)
          + 2\varphi_m\cos^2\theta\left(\frac{\varphi_m^{\prime}}{r} - \frac{\varphi_m}{r^2}\right)  \\
         = \frac{\varphi_m^2}{r^2}\sin^2\theta  +\varphi_m^{\prime 2}\cos^2\theta\\
         \langle q \rangle = \frac{\mu_1}{2}\left(\sin^2\theta\frac{\varphi_1^2 + \varphi_2^2}{r^2} + \cos^2\theta(\varphi_1^{\prime 2} + \varphi_2^{\prime 2})\right)  
        \end{gathered}
    \end{equation}
       \begin{equation}\label {B5}
        Q = \int d\bm{r} \langle q \rangle = 
        \frac{2\pi}{3}\int\limits_0^{r_0}\mu_1\left(\varphi_1^2 + \varphi_2^2 + 2\cdot r^2 
        \left\{ \left(\frac{d\varphi_1}{dr}\right)^2 + \left(\frac{d\varphi_2}{dr}\right)^2\right\}\right)dr.
    \end{equation}
\nocite{}
\bibliographystyle{unsrt}
\bibliography{bibliography}

\end{document}